\documentclass[12pt]{article}
\usepackage{graphicx,setspace,lscape,longtable}
\usepackage{epsfig,graphicx,float}
\usepackage{natbib}
\usepackage{amsmath,amsthm,amssymb,color,bm}
\usepackage[margin=1in]{geometry}
\usepackage[bottom]{footmisc}
\usepackage{indentfirst}
\usepackage{endnotes}
\usepackage{verbatim,thmtools}
\usepackage{thm-restate}
\usepackage{threeparttable}
\usepackage{comment}
\usepackage{algorithm}
\usepackage{titlesec}
\usepackage{algpseudocode}
\usepackage{multirow}
\usepackage{pdfpages}
\usepackage{dsfont}
\usepackage{subcaption}
\usepackage{xcolor}
\usepackage{rotating}
\usepackage{graphicx,wrapfig,lipsum}
\usepackage[hidelinks]{hyperref}
\usepackage{enumitem}
\RequirePackage[mathlines]{lineno}
\usepackage{multirow}
\usepackage[titletoc]{appendix}
\bibpunct{(}{)}{;}{a}{,}{,}
\floatstyle{ruled}
\newfloat{algorithm}{tbhp}{loa}
\floatname{algorithm}{Algorithm}

\newcommand{\by}{\mathrm{\bf y}}

\newcommand{\bZ}{\mathrm{\bf Z}}

\newcommand{\bD}{\mathrm{\bf D}}

\newcommand{\bI}{\mathrm{\bf I}}
\newcommand{\bK}{\mathrm{\bf K}}

\newcommand{\bO}{\mathrm{\bf O}}

\newcommand{\bX}{\mathrm{\bf X}}
\newcommand{\bY}{\mathrm{\bf Y}}

\newcommand{\bbeta}{\boldsymbol{\beta}}

\newcommand{\bone}{\mathrm{\bf 1}}
\newcommand{\balpha}{\mbox{\boldmath $\alpha$}}

\newcommand{\bzero}{\mathrm{\bf 0}}

\newcommand{\beps}{\mbox{\boldmath $\varepsilon$}}
\newcommand{\bmu}{\mbox{\boldmath $\mu$}}

\newcommand{\hmu}{\widehat\bmu}

\newcommand{\beq}{\begin{eqnarray*}}
\newcommand{\eeq}{\end{eqnarray*}}

\newcommand{\PP}{\mathbb{P}}

\def \corr {\mathrm{corr}}
\def \calD{\mbox{$\mathcal{D}$}}

\def \logit {\mathrm{logit}}

\titleformat{\section}{\normalfont\Large\bfseries}{\thesection}{0.5em}{}
\titlespacing*{\section} {0pt}{5pt}{3pt}
\titlespacing*{\subsection} {0pt}{5pt}{2pt}
\numberwithin{equation}{section}
\theoremstyle{plain}

\newtheorem{prop}{Proposition}[section]

\theoremstyle{definition}

\def\ben{\begin{equation*}}
\def\een{\end{equation*}}
\def\bea{\begin{eqnarray}}
\def\eea{\end{eqnarray}}
\def\bean{\begin{eqnarray*}}
\def\eean{\end{eqnarray*}}
\def\bep{\begin{prop}}
\def\eep{\end{prop}}
\def\bc{\begin{center}}
\def\ec{\end{center}}

\def \corr {\mathrm{corr}}
\def \calD{\mbox{$\mathcal{D}$}}

\def \logit {\mathrm{logit}}

\numberwithin{equation}{section}
\newtheorem{theorem}{\bf Theorem}

\newcommand{\blind}{1}
\begin{document}
\setstretch{1.35}
% \title{A Unifying Dependent Combination Framework with Applications to Association Tests}
\title{A Unified Combination Framework for Dependent Tests with Applications to Microbiome Association Studies}

\date{\today}

	\if1\blind
    {
    \author{ Xiufan Yu$^1$, Linjun Zhang$^2$, Arun Srinivasan$^3$, Min-ge Xie$^2$, and Lingzhou Xue$^4$ \\   $^1$University of Notre Dame, $^2$Rutgers University,  \\  $^3$GSK plc, and $^4$Penn State University }
    \date{}
    } \fi

    \if0\blind
    {
    \author{}
    } \fi

\maketitle{} 
\pagestyle{plain}

\vspace{-5ex}
\begin{abstract}
We introduce a novel meta-analysis framework 
to combine dependent tests under a general setting, and utilize it to synthesize various microbiome association tests that are calculated from the same dataset. Our development builds upon the classical meta-analysis methods of aggregating $p$-values and also a more recent general method of combining confidence distributions,
%from independent sources, 
but makes generalizations to handle 
dependent tests. 
The proposed framework ensures rigorous statistical guarantees, and we provide a comprehensive study and compare it with various existing dependent combination methods. Notably, {we demonstrate that the widely used Cauchy combination method for dependent tests, referred to as the {\it vanilla Cauchy combination} in this article, can be viewed as a special case within our framework.} Moreover, the proposed framework provides a way to address the problem when the distributional assumptions underlying the vanilla Cauchy combination are violated. Our numerical results demonstrate that ignoring the dependence among the to-be-combined components may lead to a severe size distortion phenomenon. Compared to the existing $p$-value combination methods, including the {vanilla} Cauchy combination method, the proposed combination framework can handle the dependence accurately and utilizes the information efficiently to construct tests with accurate size and enhanced power. 
The development is applied to Microbiome Association Studies, where we aggregate information from multiple existing tests {using the same dataset}. The combined tests harness the strengths of each individual test across a wide range of alternative spaces, %resulting in a significant enhancement of testing power across a wide range of alternative spaces, 
enabling more efficient and meaningful discoveries of vital microbiome associations. 
\end{abstract}

Microbiome association studies.

\newpage
%\setstretch{1}
%\doublespacing
\setstretch{1.5}
%%%%%%%%%%%%%%%%%%%%%%%%%%%%%%%%%%%%%%%%%%%%%%%%%%%%%%%%%%%%%%%%%%
\section{Introduction} \label{sec: introduction}
%%%%%%%%%%%%%%%%%%%%%%%%%%%%%%%%%%%%%%%%%%%%%%%%%%%%%%%%%%%%%%%%%%

The advance of next-generation sequencing technologies has enabled a measurement of microbiome composition by direct DNA sequencing. Detecting the associations between microbiome compositions and certain outcomes of interest (e.g., common diseases, complex traits) has received considerable attention in large-scale microbiome studies. They include associations between human microbiome and a variety of human disease and health conditions such as obesity, % \citep{turnbaugh2009},
type II diabetes, 
%\citep{qin2012},
inflammatory bowel diseases, %\citep{morgan2015associations}, 
%skin diseases, 
% \citep{schneider2019},
among others. 

Various methods for microbiome association analysis have been proposed in the literature \citep{wilson2021mirkat}. These methods behave distinctively under different types of alternatives driven by the complex nature of microbiome data, such as phylogenetic relationships among taxa, abundance information, and sparsity levels of signals. Each method possesses its own advantages and disadvantages. 
%On the one hand, 
For instance, \cite{zhao2015testing} studied distance-based association tests and showed that different choices of distance metrics largely impact the power performance across various patterns of the underlying associations, and \cite{zhan2017small,zhan2018small} extended \cite{zhao2015testing} for testing association between multiple outcomes and microbiome composition. On the other hand, \cite{koh2020powerful} further showed such distance-based tests suit the scenarios when the overall group association is large yet may suffer from power loss when the underlying signal is highly sparse. 
In real-world applications where prior knowledge about the true underlying pattern is lacking, it is important to derive a robust testing procedure that remains powerful against general alternatives.

A natural procedure is to take advantage of existing statistics and construct a new test by combining the $p$-values, such as Fisher's combined probability tests \citep{yu2020fisher,yu2019innovated} and the adaptive tests \citep{xu2016adaptive,he2021asymptotically}. 
Both approaches require independence among the to-be-combined components. 
There is a rich literature on meta-analysis for combining various studies in situations when the to-be-combined components are independent; see \cite{borenstein2021introduction} for a summary. \cite{singh2005combining} and \cite{xie2011confidence} developed a unifying framework for combining information from independent studies using confidence distributions. 
The framework unifies classical methods of combining $p$-values and modern model-based meta-analysis approaches, providing a general approach for aggregating information from independent studies.

However, the assumption of independence is quite restrictive in practice. Especially when the components are constructed based on the same dataset, the assumption of independence 
is often violated. How to combine dependent tests is a long-standing open question in statistics (cf., \href{en.wikipedia.org/wiki/List_of_unsolved_problems_in_statistics}{wikipedia.org: ``List of unsolved problems in statistics''}) until recently, and more research works remain to be developed. Specifically, \cite{liu2020Cauchy} recently proposed a new and popular approach for dependent combination (referred to as {\it the vanilla Cauchy combination} in our article), which suggests to construct a test statistic $T_{Cauchy}$ taking the form of a weighted sum of Cauchy transformation of individual $p$-values. 
If the individual
$p$-values are independent or perfectly dependent, $T_{Cauchy}$ is naturally a standard Cauchy variable. For other dependent cases, \cite{liu2020Cauchy} used the result of 
\cite{pillai2016unexpected} to
prove that the tail behavior of $T_{Cauchy}$ could be reasonably approximated by a Cauchy distribution, if the marginal statistics follow a bivariate normal distribution with any correlation structure. Such bivariate normal assumption accommodates the dependency to some extent but does not cover  arbitrary joint distributions. The tail distribution of $T_{Cauchy}$ remains an open question for arbitrarily distributed test statistics. Additionally, the Cauchy approximation is only valid for approximating tail probability $P(T_{Cauchy}>t)$ when $t \rightarrow \infty$. 
Such tail approximation provides a useful tool for hypothesis testing problems when people pay the most attention to small $p$-values. Nevertheless, it fails to provide a full picture as the (asymptotic) distribution of  $T_{Cauchy}$ remains unknown. Several follow-up works \citep[e.g.,][] {fang2023heavy,long2023cauchy} are developed, but the issues remain. 

To further explore the Cauchy combination method, we conduct extensive simulations to study the test performance when violating the bivariate normal assumption. On the one hand, some numerical results suggest the vanilla Cauchy combination may perform reasonably well even when the assumptions are not fully satisfied. For example, in Section \ref{subsec: simulation-microbiome}, we observe that the {vanilla} Cauchy 
method 
(without using any adjustment for dependence) exhibits comparable performance to the dependence-adjusted Cauchy (to be proposed) in terms of empirical size and power. On the other hand, we also discover 
that 
applying the vanilla Cauchy combination blindly without any correction for the dependence will lead to size inflation. More specifically, Section \ref{subsec: simulation-dCauchy} shows a synthetic setting in which the vanilla Cauchy combination fails to control the type-I error, whereas {our} dependence-adjusted Cauchy retains the correct size after accommodating the dependence. 
{It is clear that, based on the empirical evidence, the validity of the {vanilla} Cauchy combination}
% its validity 
cannot be guaranteed in cases with arbitrary distributional assumptions and arbitrarily dependent structures. 

In this paper, we propose a novel general framework to combine arbitrarily dependent $p$-values with statistical guarantees under arbitrary distributional assumptions.
The main contributions can be summarized in four aspects.

\emph{First}, we extend the general recipe of \cite{xie2011confidence} for combining independent studies to allow for dependence among those to-be-combined components. Our proposed framework does not impose any distributional assumptions on joint distributions and is applicable in combining arbitrarily dependent components. 
Under the framework and tailored to the aforementioned application of microbiome association studies, we introduce a parametric-bootstrap-based method 
to recover and accommodate the unknown dependence among the components {obtained using the same dataset}.
Our theoretical studies prove that the procedure produces legitimate combined $p$-values that achieve guaranteed accurate size. 

\emph{Second}, we connect {the proposed framework} to various independent and dependent combination methods. We show the proposed framework coincides with many conventional meta-analysis methods (e.g., Fisher's method \citep{Fisher1925}, 
Stouffer's method \citep{stouffer1949adjustment})
under independent cases, while it provides proper adjustments to account for potential dependence when handling dependent tests. 
Moreover, the framework subsumes the {recently developed dependent combination methods, including the vanilla Cauchy combination and also the closely related works of \cite{wilson2019harmonic} and \cite{fang2023heavy}. }

\emph{Third}, the practical effectiveness of the proposed framework is demonstrated using an application to microbiome association tests. 
By aggregating the information from various tests that possess their own advantages under different types of alternatives, the combined test retains the advantages of each test by inheriting their respective high-power regions, and hence the power is enhanced towards a wide range of alternative spaces. 
The proposed framework provides efficient tools to construct a dependence-adjusted combined test that is powerful and robust to various alternative patterns.

\emph{Fourth}, we provide {a comparative study to examine} 
various combination methods and discuss their respective efficiency. \cite{singh2005combining} provided discussions on the efficiency 
for combining independent information. We generalize the discussions and use bivariate normal tests to examine the performance of various combination methods under our framework in situations of arbitrary dependence. We show  Stouffer's method {is the most powerful for the dependent normal tests} while the vanilla Cauchy combination and other methods will loss some power under the same setting.
This efficiency message for combining dependent tests is the same as that of \cite{xie2011confidence} for combining independent $p$-values.

The paper is organized as follows. Section \ref{sec: methodology} introduces methodological details of the proposed unifying framework for combining dependent $p$-values.
Section \ref{sec: association-tests} is an illustrative example of using the proposed framework to construct new association tests for microbiome studies. 
Section \ref{sec: simulation} conducts simulation studies to illustrate the necessity of adjusting dependence in the combination and demonstrate the efficacy of our proposed combining framework. 
Section \ref{sec: realdata} further shows its effectiveness using two real data examples.
Section \ref{sec: various-combination-methods} discusses the efficiency of various combination methods. 
Section \ref{sec: conclusion} includes a few concluding remarks. Technical proofs are provided in the appendices.

%\newpage
%%%%%%%%%%%%%%%%%%%%%%%%%%%%%%%%%%%%%%%%%%%%%%%%%%%%%%%%%%%%%%%%%%
%%\vspace{-5ex}
%\section{A Unifying Dependent Combination Framework} \label{sec: methodology}
\section{Methodology}\label{sec: methodology}
%%%%%%%%%%%%%%%%%%%%%%%%%%%%%%%%%%%%%%%%%%%%%%%%%%%%%%%%%%%%%%%%%%
%In this section, we first review existing works on combining independent studies in Section \ref{subsec:method-indep-combination}. Then in Section \ref{subsec:method-dep-combination}, we propose a more general framework that is able to accommodate dependence in the combination. 
% ############################################################## %
\subsection{Revisiting the Unifying Framework for Combining Independent Studies} \label{subsec:method-indep-combination}
% ############################################################## %

Aggregating information from multiple aspects has a long-standing history in statistics.  In the scientific community, 
it serves as a vital tool for researchers to combine multiple scientific studies and derive conclusions on a particular topic of interest across a broad range of disciplines, e.g., ecology, % \citep{gurevitch2018meta}, 
%medical research, %\citep{claggett2014meta}, 
psychology, % \citep{bolier2013positive}, 
and social science. %Utilize the combination of information from multiple aspects
It also plays an essential role in many modern statistical methods to improve algorithm performance, such as divide-and-conquer approaches \citep{chen2014split, wang2021multivariate}, data-splitting-based approaches \citep{chen2018error,cui2018test,liu2021multiple}, and power enhancement tests \citep{fan2015power,yu2022power,yu2019innovated,cammarata2022power}. 

\cite{xie2011confidence} proposed a generic framework for combining studies assuming the studies are independent. The framework unifies classical methods of combining independent $p$-values.
Specifically, consider a hypothesis of interest $H_0$, and let %$p_i = p_i(\calD_i)$
{$p_i = p_i^{\calD_i}$} be the $p$-value for testing $H_0$ from the $i$-th study with sample dataset $\calD_i$ and its realization $\calD_{i, \rm{obs}}$, for $i=1,\cdots, k$. Assume that the $k$ studies are independent (i.e., $\calD_1, \ldots, \calD_k$ are independent), the generic framework of \cite{xie2011confidence} suggests to combine the $k$ studies as follows:
%\vspace{-2ex}
\begin{equation}\label{eq:ind}
    p_c = G_c(g_c(p_1, \cdots, p_k)).
    %\vspace{-2ex}
\end{equation}
The function $g_c(u_1,\cdots, u_k)$ is any continuous function from the $k$-dimension hybercube $[0,1]^k$ to the real line $\mathbb{R} = (-\infty, \infty)$ which is monotonic with respect to each coordinate. The function $G_c(\cdot)$ is the cumulative distribution function (CDF) for $g_c(U_1,\ldots,U_k)$, i.e.
$
G_c (s) = \PP\left(g_c(U_1,\ldots, U_k)\leq s\right),
$
where $U_1, \cdots, U_k$ are i.i.d. $\mathcal{U}[0,1]$ random variables.

Many classical $p$-value combining approaches can be regarded as special cases of this unifying framework (\ref{eq:ind}); see Table \ref{tab: gcfuns} for a list of commonly used combination methods and their corresponding $g_c$ functions. 
For example, take $g_c$ of the form
$g_c(u_1, \cdots, u_k) = \log (u_1) +\ldots + \log (u_k)$,
then $G_c(s)$ becomes
$G_c(s) = \PP\left(g_c(U_1,\cdots, U_k)\leq s\right) = \PP(\sum_{i=1}^k \log (U_i) \leq s) = \PP(\chi_{2k}^2 \geq -2s)$ 
for $U_1,\cdots, U_k \overset{i.i.d.}\sim \mathcal{U}[0,1]$, so that the resulting $p$-value obtained by
$p_c = G_c(g_c(p_1, \cdots, p_k)) = \PP(\chi^2_{2k} \geq -2 \sum_{i=1}^k \log {p_i})$ is exactly the Fisher's method. 

%\vspace{-2ex}
\begin{table}
\captionsetup{justification=centering}
\caption{Combination methods and their corresponding $g_c$ functions 
% \\ (with the assumption that $U_1, \ldots, U_k$ are independent.) 
}\label{tab: gcfuns}

\centering
%%\vspace{-2ex}
%\small
\begin{tabular}{cclllllllll}
\hline
Method & Notation & $g_c$ function \\ \hline
Fisher & Fisher & $g_c(u_1, \ldots, u_k) = \log (u_1) + \ldots + \log (u_k)$\\ \hline
Stouffer & Stou & $g_c(u_1, \ldots , u_k) = \Phi^{-1}(u_1) + \ldots +  \Phi^{-1}(u_k)$ \\ \hline
Double Exponential  & DE & $g_c(u_1, \ldots, u_k) = F_0^{-1}(u_1) + \ldots + F_0^{-1}(u_k)$ \\
&& where $F_0(t) = 0.5 e^{t} \bone\{t \leq 0\} + (1-0.5 e^{-t}) \bone\{t > 0\}$  \\ \hline
Minimum & Gcmin & $g_c(u_1, \ldots,u_k) = \min\{u_1,u_2, \ldots, u_k \}$ \\ \hline
Cauchy  & Cauchy & $g_c(u_1, \ldots, u_k) = w_1F_{CC}^{-1}(u_1) + \ldots + w_k F_{CC}^{-1}(u_k)$ \\
&& where $F^{-1}_{CC}(u) = \tan [(u - 0.5) \pi]$ is the inverse CDF of \\
&& the standard Cauchy distribution, $\sum_{i=1}^k w_i = 1$, $w_i \geq 0$. \\ 
\hline
Harmonic Mean & HM & $g_c(u_1, \ldots,u_k) = u_1^{-1}+\ldots + u_k^{-1}$ \\  \hline
Pareto & Pareto & $g_c(u_1, \ldots,u_k) = u_1^{-\eta}+\ldots + u_k^{-\eta}$ for $\eta>0$ \\ \hline
\end{tabular}
\end{table}
It is worth mentioning that  
the vanilla Cauchy combination by \cite{liu2020Cauchy} is a special case of (\ref{eq:ind}) with $g_c(u_1, \ldots, u_k) = w_1F_{CC}^{-1}(u_1) + \ldots + w_k F_{CC}^{-1}(u_k)$. Here, $F^{-1}_{CC}(u) = \tan [(u - 0.5) \pi]$ is the inverse CDF of the standard Cauchy distribution and $\sum_{i=1}^k w_i = 1$ for given weights $w_i \geq 0$.
In appearance, the Cauchy combination in Table \ref{tab: gcfuns} may seem to be different
than that in \cite{liu2020Cauchy},
but the two definitions are in fact equivalent.
   In particular,  \cite{liu2020Cauchy} defined the test statistic as $T_{CC}(u_1,\ldots, u_k) = \sum_{i=1}^k w_i\tan[(0.5-u_i)\pi]$, which equals to $-g_c(u_1,\cdots,u_k)$.  The associated combined $p$-value is obtained by $p_{LX} = 1-F_{CC}(T_{CC}(u_1,\ldots, u_k))$. When $U_1,\ldots,U_k$ are independent, by definition, we know $G_c(s) = F_{CC}(s)$ and the $p$-value in (\ref{eq:ind}) becomes $p_{c} = G_c(g_c(u_1, \ldots, u_k)) = F_{CC}(g_c(u_1, \ldots, u_k))$. Hence, the two definitions are equivalent as the combined p-value by \cite{liu2020Cauchy} is 
   $p_{LX} = 1-F_{CC}(T_{CC}(u_1,\ldots, u_k)) = 1-F_{CC}(-g_c(u_1, \ldots, u_k))) = 1-[1-F_{CC}(g_c(u_1, \ldots, u_k))] = F_{CC}(g_c(u_1, \ldots, u_k)) =  p_{c}$, the combined p-value by (\ref{eq:ind}). 

Several other recently proposed dependent combination methods are also special cases of (\ref{eq:ind}). For instance, the harmonic mean combination method by \cite{wilson2019harmonic},  taking the form of $p_{HM}=(\sum_{i=1}^k \frac{1}{p_i})^{-1}$, is also a special case of (\ref{eq:ind}). In addition, \cite{fang2023heavy} 
extended the works of \cite{liu2020Cauchy} and \cite{wilson2019harmonic} to a broad family of heavy-tailed distributions, e.g., the Pareto distributions. 
The test statistic is defined as $T_{Pareto} = \sum_{i=1}^k F_{Pareto(\eta^{-1},1)}^{-1}(1-p_i) = \sum_{i=1}^k p_i^{-\eta}$ for some $\eta>0$, where $F_{Pareto(\eta^{-1},1)}^{-1}(\cdot) $ is the inverse CDF of the $\rm{Pareto}(1/\eta,1)$ distribution. This Pareto-transformed combination framework unifies the vanilla Cauchy combination, harmonic mean combination, and a family of combination methods utilizing heavy-tailed distributions. It is also a special case of the unifying framework (\ref{eq:ind}).

% ############################################################## %
%\vspace{-2ex}
\subsection{A Unifying Dependent Combination Framework} \label{subsec:method-dep-combination}
% ############################################################## %
The independence among the $p$-values $\{p_1, \ldots, p_k\}$ is essential to guarantee the validity of the resultant {$p$-value $p_c$ of (\ref{eq:ind})}. 
Numerical studies in Section \ref{sec: simulation} show that when the independence assumption is violated, applying the above methods blindly without any correction for the dependence will lead to severely inflated type-I errors.  Such independence holds naturally in a meta-analysis when the $p$-values are from independent sources. However, 
in many cases, it is not reasonable to assume independence. For example, {in our application of combining microbiome association tests, the tests are computed from the same data and thus dependent.}
{To better handle the dependence, we propose to} extend the framework (\ref{eq:ind}) to combining dependent $p$-values. Specifically, we define
\begin{equation*}
    \widetilde{G}_c (s) = \PP\left(g_c(U_1,\ldots, U_k)\leq s\right).
    %\vspace{-1.5ex}
\end{equation*}
in which $(U_1, \cdots, U_k)$ are marginally $\mathcal{U}[0,1]$ but they can be arbitrarily dependent. The dependence-adjusted combined $p$-value is then defined as 
%\vspace{-1ex}
\begin{equation}\label{eq: tildepc}
    \widetilde p_c = \widetilde G_c\left( g_c( p_1^{\calD_1}, \ldots, p_k^{\calD_k}) \right),
    %\vspace{-1.5ex}
\end{equation}
where {$p_i^{\calD_i}$} is the $p$-value for testing $H_0$ from the $i$-th study with sample dataset $\calD_i$ and its realization $\calD_{i, \rm{obs}}$. Here, $\mathcal{D}_1, \ldots, \mathcal{D}_k$ may be dependent or even be the same dataset (i.e., $\calD_1 \equiv \ldots \equiv \calD_k \equiv \calD$). The $g_c(\cdot)$ function is the same as that used in (\ref{eq:ind}), listed in Table \ref{tab: gcfuns}. It follows immediately that $\widetilde p_c  \sim \mathcal{U}[0,1]$ under the hull hypothesis $H_0$.
This $\widetilde p_c$ is a combined $p$-value that incorporates information from the $k$ dependent $p$-values $p_1^{\calD_1}, \ldots, p_k^{\calD_k}$.

Note that, to distinguish the independence-assumed combination approaches \citep{xie2011confidence} and the proposed dependence-adjusted combination methods, here we have used two sets of different notations for the two frameworks:
% . That is, 
$(G_c(\cdot), p_c)$ denote the CDF and the combined $p$-value from the independence-assumed combination, and $(\widetilde{G}_c(\cdot), \widetilde{p}_c)$ denote those from the proposed dependence-adjusted combination. The lower case $g_c(\cdot)$ function that facilitates our combination method is the same for the two frameworks, 
regardless whether or not the $p$-values are dependent.

%%%%%%%%%%%%%%%%%%%%%%%%%%%%%%%%%%%%%%%%%%%%%%%%%%%%%%%%%%%%%%%%

The vanilla Cauchy combination can be regarded as a special case of the independent combination framework \eqref{eq:ind} but not the proposed framework \eqref{eq: tildepc}. When the to-be-combined $p$-values $p_1, \cdots, p_k$ are independent, the Cauchy output from \eqref{eq:ind}
(with $g_c$ taking the form shown in Table \ref{tab: gcfuns})
provides a valid $p$-value in the sense that it follows $\mathcal{U}[0,1]$ under the null. However, when $p_1, \cdots, p_k$ are dependent, the Cauchy output is no longer a valid $p$-value. \cite{liu2020Cauchy} proved that when assuming $p_1, \cdots, p_k$ are obtained from bivariate normal variables, the combined output can be approximated by the tail probability of a standard Cauchy variable. Such approximation fundamentally differs from the proposed dependent combination framework \eqref{eq: tildepc}: the vanilla Cauchy combination provides a numerical approximation but the combined output itself does not follow $\mathcal{U}[0,1]$ under $H_0$ when $p_1, \ldots, p_k$ are dependent, while the framework \eqref{eq: tildepc} explicitly models the dependence among $p_1, \ldots, p_k$ and therefore produces a valid $p$-value under arbitrary dependence. In addition, we would like to mention % bring up a reminder 
that, to our best knowledge, % by far, 
the theoretical justification of Cauchy approximation is established only under the assumptions that the statistics are bivariate normal variables. When this assumption is violated, there is no guarantee that such an approximation is still valid. Though the vanilla Cauchy combination exhibits appealing characteristics of insensitiveness to dependence among statistics, it cannot control the type-I error exactly under arbitrarily dependent situations. In Section \ref{subsec: simulation-dCauchy}, we study a synthetic setting in which the vanilla Cauchy combined test fails to control the type-I error rate. Hence adjustments to account for the dependence are desired when the assumptions are violated. 
%\end{remark}

%%%%%%%%%%%%%%%%%%%%%%%%%%%%%%%%%%%%%%%%%
\vspace{-1ex}
\subsection{Empirical Characterization of the Dependence}
%%%%%%%%%%%%%%%%%%%%%%%%%%%%%%%%%%%%%%%%%
{In the proposed dependent combination framework, the dependence is incorporated}
through the $\widetilde{G}_c(\cdot)$ function, defined as $\widetilde{G}_c (s) = \PP\left(g_c(U_1,\ldots, U_k)\leq s\right)$ in which $(U_1, \cdots, U_k)$ are marginally $\mathcal{U}[0,1]$ but can be arbitrarily dependent. In general, the dependence among the $p$-values $p_1, \ldots, p_k$ (and therefore the dependence among $(U_1, \ldots, U_k)$ in the $\widetilde{G}_c (\cdot)$ function) may not be known. Thus, the explicit form of $\widetilde G_c (\cdot)$ is not readily available in practice.

\newcounter{myalgo}
\setcounter{myalgo}{1} 

{We propose to use} a bootstrap sampling method to approximate $\widetilde G_c(\cdot)$ by a bootstrap empirical CDF $\widetilde G^B_c(\cdot)$.
This bootstrap approach can be directly applied to the application of microbiome association studies to combine various tests conducted using the same data set and harness the strengths of each individual test across a wide range of alternative spaces.  
Algorithm \themyalgo\ (in Table \ref{algo: combining-dependent-$p$-values}) summarizes the procedures of combining dependent $p$-values, where we preserve the dependence across multiple tests. Theorem \ref{thm: bootstrappedPvalue} below provides theoretical guarantees for using the bootstrapped estimators (\ref{eq: pcB}) to calculate the combined $p$-value (\ref{eq: tildepc}), and %. Theorem \ref{thm: bootstrappedPvalue} 
it proves the validity of employing (\ref{eq: pcB}) for hypothesis tests.

%https://www.overleaf.com/learn/latex/Counters#\setcounter{somecounter}{number}

%\vspace{-2ex}
\begin{table}
%\begin{algorithm}
\caption{The proposed algorithms}
\begin{subtable}{\textwidth}
\captionsetup{justification=centering}
%\caption{\normalsize Algorithm 1\ -- A Unifying Framework for Combining Dependent $p$-values}\label{algo: combining-dependent-$p$-values}
\caption{Algorithm 1\ -- A Unifying Framework for Combining Dependent $p$-values}\label{algo: combining-dependent-$p$-values}
\noindent\rule{\textwidth}{0.4pt}
\begin{algorithmic}[1] 
\State Use parametric bootstrapping method to construct the bootstrap samples $\mathcal{D}_1^b, \cdots, \mathcal{D}_k^b$.
\State Calculate the corresponding $p$-value $p_{1}^{b} = p_{1}^{b, \mathcal{D}_1^b}, \ldots, p_{k}^{b} = p_{k}^{b, \mathcal{D}_k^b}$.
\State Approximate $\widetilde{G}_c(z)$ by the empirical CDF
%\begin{equation*}
    $\widetilde{G}_c^B(z) = \frac{1}{B}\sum_{b=1}^B \bone\{g_c(p_1^b, \cdots , p_k^b) \leq z\}.$
%\end{equation*}
\State Then the combined $p$-value can be represented by
\begin{equation} \label{eq: pcB}
    \widetilde p_c^B = \widetilde{G}_c^B(g_c(p_1, \cdots , p_k)).
\end{equation}
where $p_1, \cdots, p_k$ are the $p$-value from the original sample $\mathcal{D}_1, \ldots, \mathcal{D}_k$.
\end{algorithmic}
\noindent\rule{\textwidth}{0.4pt}
%\end{algorithm}
\end{subtable}

\bigskip\bigskip
\begin{subtable}{\textwidth}
\captionsetup{justification=centering}
%\caption{\normalsize Algorithm 2\ -- Combining Dependent $p$-values of MiRKAT and MiHC tests} \label{algo: MiRKAT-MiHC}
\caption{Algorithm 2\ -- Combining Dependent $p$-values of MiRKAT and MiHC tests} \label{algo: MiRKAT-MiHC}
\noindent\rule{\textwidth}{0.4pt}
\begin{algorithmic}[1]
%\State Use model-based bootstrapping method, bootstrap $\beps^b$, $b = 1, \cdots, B$, from the residuals $\beps$ fitted by original sample $\left(\bX, \bG, \bY\right)$, and construct the bootstrap samples as $\left(\bX^b, \bG^b, \bY^b\right)=\left(\bX, \bG, \widehat{\bY}+\beps^b\right)$
\State  \emph{If $\bY$ is a continuous phenotype:} Regress $\bY$ on $\bX$ and estimate the variance of $\varepsilon$, denoted by $\widehat\sigma^2$. Use parametric bootstrapping method to bootstrap $\beps^b$, $b = 1, \cdots, B$, from $N_n(0, \widehat\sigma^2\bI_n)$, and construct the bootstrap samples as $\left(\bX^b, \bZ^b, \bY^b\right) \leftarrow (\bX, \bZ, \widehat{\bY}+\beps^b)$. 

\noindent \emph{If $\bY$ is a dichotomous phenotype:}
    Estimate the coefficient $\widehat\balpha$ in the logistic regression under the null hypothesis $\bbeta = \bzero$. Construct the bootstrap sample as $\left(\bX^b, \bZ^b, \bY^b\right) \leftarrow \left(\bX, \bZ, (Y_1^b, \cdots, Y_n^b)' \right)$, where 
$Y_i^b\sim Ber(\phi(\widehat\balpha'\bX_i))$ and $\phi(x)=(1+e^x)^{-1}$. 
\State Calculate $Q_{MiRKAT}^{b}$ and $M_{MiHC}^{b}$ and their corresponding $p$-value $p_{Q}^{b}$ and $p_{M}^{b}$.
\State Approximate $\widetilde{G}_c(z)$ by the empirical CDF
$\widetilde{G}_c^B(z) = \frac{1}{B}\sum_{b=1}^B \bone\{g_c(p_{Q}^b, p_{M}^b) \leq z\}.$
\State Then the combined $p$-value can be represented by
$\widetilde p_c^B = \widetilde{G}_c^B(g_c(p_Q, p_M)),$
where $p_Q$ and $p_M$ are the $p$-value of MiRKAT and MiHC from the original sample $(\bX, \bZ, \bY)$.
\State The test rejects $H_0^G$ at the significance level $\alpha$ if 
    $\widetilde p_c^B \leq \alpha.$
\end{algorithmic}
\noindent\rule{\textwidth}{0.4pt}
\end{subtable}
\end{table}
%\vspace{-2ex}

\vspace{-1ex}
\begin{theorem} \label{thm: bootstrappedPvalue}
%{\bf Theorem 1} 
Suppose 
{ 
$(p_1^{b, \calD_1^b}, \ldots , p_k^{b, \calD_k^b})\stackrel {\mathcal L}{\to} (p_1^{\calD_1}, \ldots, p_k^{\calD_k})$,
} as $n \to \infty$.
Given a smooth and coordinate-wise monotonic function $g_c$ from $[0,1]^k \to {\mathbb{R}} = (-\infty, \infty)$. Then, as $n \to \infty$,
$
\widetilde p_c^* = \widetilde{G}_c^*(g_c(p_1^{\calD_1}, \ldots, p_k^{\calD_k})) \overset{\cal L}{\to} \mathcal{U}[0,1],
$
where $\widetilde G_c^*(s) = \text{\rm P}^*( g_c(p_1^b, \ldots , p_k^b) \leq s \mid   {\calD}_{\rm obs})$, 
for $s \in (-\infty, \infty)$, and 
{
$p_1^b = p_1^{b, \calD_1^b}$, $\ldots$, $p_k^b = p_k^{b, \calD_k^b}$
}
are the $p$-values computed using a random bootstrap sample. Furthermore, if $\widetilde G_c^B(s)$ is the empirical estimate of $\widetilde G_c^*(s)$ obtained from $B$-copies of bootstrap samples, 
then, as $n, B \to \infty$, 
{ 
$
%%\vspace{-2ex}
\widetilde p_c^B = \widetilde{G}_c^B(g_c(p_1^{\calD_1}, \ldots, p_k^{\calD_k})) \overset{\cal L}{\to} \mathcal{U}[0,1].  
$
}
\end{theorem}

%A proof of the theorem is in Appendix \ref{supp-sec: lemmas-proofs}.
A proof of the theorem is in Appendix.
 Alternatively, one may model joint dependence using copulas. 
\cite{long2023cauchy} studies the validity of Cauchy approximation when certain assumptions are imposed on pairs of $p$-values rather than test statistics, and shows these assumptions are satisfied by six popular copula distributions. We would like to clarify that in the same vein as the discussions in Section \ref{subsec:method-dep-combination}, 
%Remark \ref{remark:Cauchy},
the approximation of tail probability proposed in \cite{long2023cauchy} is fundamentally different from our proposed dependent combination framework \eqref{eq: tildepc}. The tail approximation in \cite{long2023cauchy} only works under certain conditions (imposed either on pairs of test statistics or pairs of $p$-values) and it fails to provide a valid $p$-value in situations of arbitrary dependence.

%%%%%%%%%%%%%%%%%%%%%%%%%%%%%%%%%%%%%%%%%%%%%%%%%%%%%%%%%%%%%%%%%%
%\vspace{-3ex}
%\section{Dependence-Adjusted Combined Association Tests in Genetic Studies} \label{sec: association-tests}
%\section{Dependence-Adjusted Combined Association Tests in Microbiome Studies} \label{sec: association-tests}
\section{An Application to Microbiome Association Tests} \label{sec: association-tests}
%%%%%%%%%%%%%%%%%%%%%%%%%%%%%%%%%%%%%%%%%%%%%%%%%%%%%%%%%%%%%%%%%%

%This section studies an illustrative example of utilizing our proposed dependent combination framework to construct new association tests for microbiome association studies.

% ############################################################## %
\subsection{Problem Setup}\label{subsec: microbiome-problem-setup}
% ############################################################## %
Human microbiome is the community of microbes in and around the human body. A key public health question is to better understand the relationship between these microbes and the development of disease. Association testing plays a key role in studying the links between microbes and disease pathogenesis. 
Microbiome data consists of the counts of the number of DNA strands associated with a given microbe of interest, known as operational taxonomic units (OTU), in each sample. OTUs serve as proxies for the microbes that are within a sample, and a larger OTU count for a given microbe indicates a higher abundance of the microbe within the sample. Thus, a common question in microbiome analysis is determining if a set of OTU features is related to a particular outcome such as disease status.

We consider a study of $n$ subjects with $p$ OTUs (microbes) of interest. We define $y_i$ as the biological outcome of interest for subject $i$ and $\bY=(y_1,...,y_n)'$ as the vector of outcomes. Further consider an $n\times q$ matrix $\bX = (X_{ij})_{1\leq i \leq n, 1\leq j\leq q}$ as the matrix of covariates in which the $(i,j)$-th element $X_{ij}$ denotes the value of covariate $j$ for subject $i$, and  an $n\times p$ matrix $\bZ = (Z_{ik})_{1 \leq i \leq n, 1\leq k \leq p}$ where $Z_{ik}$ denotes the count for OTU $k$ from subject $i$. {Let the row vector $\mathbf{Z}_i=(Z_{i1},...,Z_{ip})'$ represent the counts of OTUs for the subject $i$; thus, $\bZ = (\bZ_1,\ldots,\bZ_n)'$. 
We follow common practice by considering the relative abundance (i.e., proportion) of the OTUs. Let $\bO = (o_{ik})_{1 \leq i \leq n, 1\leq k \leq p}$ be an $n\times p$ matrix where $o_{ik} = Z_{ik}/\sum_{j=1}^p Z_{ij}$ is the proportion of OTU $k$ in subject $i$. Similar to $\bZ$, we denote $\bO =(\bO_1,\ldots,\bO_n)'$, where $\bO_i = (o_{i1}, \ldots, o_{ip})'$ is the OTU relative abundance for subject $i$. For example, in the association study between human respiratory microbiome community and smoking status (presented in Section \ref{sec: realdata}), the response $\bY$ is smoking status, the covariates $\bX$ include clinical covariates (such as gender) and medical history (e.g., whether the individual had used antibiotics in the last 3 months), the matrix $\bZ$ is the throat microbiome OTU counts, and the compositional matrix $\bO$ is the relative throat microbiome OTU  abundance that can be directly computed from the OTU counts. } Given this, we study the linear model,
%\begin{equation} \label{eq: microbiome-LM}
    $y_i = \bX_i'\balpha+\bO_i'\bbeta + \varepsilon_i$
%\end{equation}
for continuous outcomes, and the logistic regression model 
%\begin{equation} \label{eq: microbiome-LogisticModel}
$\logit\ P(y_i=1) = \bX_i'\balpha+\bO_i'\bbeta$
%\end{equation}
for binary outcomes where $\balpha = (\alpha_1,..,\alpha_q)'\in \mathbb{R}^q$ and $\bbeta=(\beta_1,...,\beta_p)'\in \mathbb{R}^p$ are the vector of regression coefficients for the $q$ covariates and $p$ OTUs. The key test of interest has the corresponding null hypothesis
%\begin{equation} \label{eq: micro-H0}
    $H_0: \bbeta = (\beta_1, \cdots, \beta_p)' = \bzero$, {which assesses the association between the outcome variable and the overall microbiome community after adjusting for the covariates. }
%\end{equation}
% ############################################################## %
%\vspace{-2ex}
\subsection{Existing Methods}
\label{subsec: microbiome-existing-methods}
% ############################################################## %
There is a vast array of literature on association tests for microbiome analysis. Among various methods, we employ two popular methods, the Microbiome Kernel Association Test (MiRKAT) \citep{zhao2015testing} and the Microbiome Higher Criticism Test (MiHC) \citep{koh2020powerful}, in our illustrative example. 
In fact, our proposed combination method is extremely flexible allowing for the use of an array of different tests.

MiRKAT is a semiparametric, kernel-based testing procedure that can be flexibly applied to a host of microbiome association testing problems. We define $\bK=(K_{ij})_{1\leq i,j \leq n}$ as the $n \times n$ kernel which measures the similarity between the $i$-th OTU  and the $j$-th OTU, that is, a function of $\bm Z$. We construct the kernel through transforming an $n \times n$ distance matrix $\bD$ by letting
%\begin{equation}\label{eq: kernel-micro}
    $\bK = -\frac{1}{2}(\bI - \frac{\mathbf{1}\mathbf{1}'}{n})\bD^2 (\bI - \frac{\mathbf{1}\mathbf{1}'}{n}),$
%\end{equation}
where $\bI$ denotes the identity matrix and $\mathbf{1}$ is a vector of ones. The choice of $\bD$ determines what information should be accounted for in measuring similarity. While there is a wide array of options for measuring distance, we will use the Bray-Curtis dissimilarity metric \citep{bray1957ordination} in our example. Given this kernel, we are able to define the following test statistic,
%%%\vspace{-2ex}
%\begin{equation}\label{test-stat: MiRKAT}
$Q_{MiRKAT} = \frac{1}{2\phi}(\bY-\hmu)'\bK(\bY-\hmu),$
%%%\vspace{-2ex}
%\end{equation}
where $\phi$ denotes the dispersion parameter. In the linear kernel regression $\phi = \widehat{\sigma}^2$ (the estimated residual variance under the null model) and $\phi=1$ for logistic kernel regression. Under $H_0$, the $Q_{MiRKAT}$ follows a mixture of weighted chi-square distributions. The $p$-value for MiRKAT, denoted by $p_Q$, can be computed analytically via approaches detailed in \cite{zhao2015testing}. 

The MiHC procedure is an adaptation of higher criticism to microbiome data analysis. The MiHC statistic is defined through a combination of modulated weighted higher criticism ($wHC(h)$), modulated unweighted higher criticism ($uHC(h)$), and the Simes test \citep{simes1986improved}. The tuning parameter $h\in\Gamma$ controls sensitivity to the sparsity level of the underlying signal. We refer to Equations (8)-(10) in \cite{koh2020powerful} for the explicit formulas of $wHC(h)$, $uHC(h)$, the Simes statistics, and the set $\Gamma$. 
Let $p_{uHC(h)}$, $p_{wHC(h)}$, $p_{\text{Simes}}$ denote the $p$-values obtained from the aforementioned three tests. The MiHC test statistic is defined as 
$M_{MiHC} = \min\left( \min_{h\in \Gamma}\left( p_{uHC(h)}, p_{wHC(h)}), p_{\text{Simes}}\right)\right),$
%%%\vspace{-2ex}
%\end{equation}
The $p$-value for MiHC, denoted by $p_M$, can be computed according to the procedures outlined in 
\cite{koh2020powerful}.

% ############################################################## %
%\vspace{-3ex}
\subsection{\large A Dependence-Adjusted Combined Microbiome Association Test}\label{subsec: microbiome-bootstrap}
% ############################################################## %
The MiRKAT and MiHC behave differently under various signal settings. 
To further illustrate it, we let $\mathcal{A}$ denote the set of indices of significant OTUs, and consider the following three common signal structures depending on how the
active signal set is defined.

%\vspace{-2ex}
\begin{enumerate}
    \item \textbf{Phylogenetic Setting:} We partition all OTU in 20 different lineages by OTU distance. We select one of the cluster lineages to be a causal cluster and all other OTUs are noise.
    \item \textbf{Abundance Setting:} We sort the OTU matrix by total OTU count and take define the top $K\%$ most abundant OTU as signal set $\mathcal{A}$ where $K$ denotes the sparsity level.
    \item \textbf{Random Setting:} We randomly choose $K\%$ of the OTU features, where $K$ denotes the sparsity level, as the signal set.
\end{enumerate}

\begin{figure}%[!hbt]
    \centering
    \includegraphics[width=0.75\textwidth]{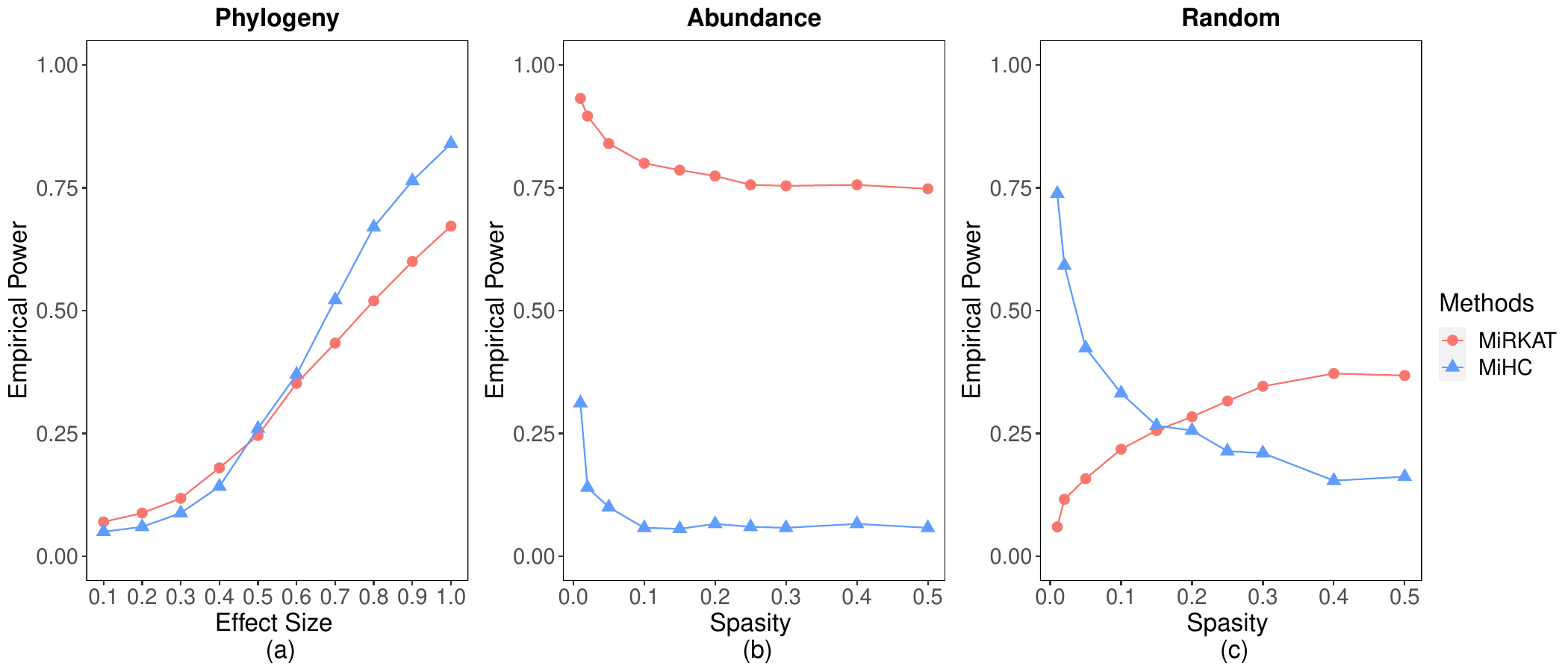}
    %%\vspace{-1ex}
    \captionsetup{justification=centering}
    \caption{Demonstration of various performances of MiRKAT and MiHC tests under different situations. Data for (a) are generated under the phylogenetic setting with $\beta$ ranging from 0.1 to 1 as a multiplier of 0.1. Data for (b) and (c) is generated under the abundance and random signal settings
    %, respectively, 
    with the sparsity level taking values in $\{0.01, 0.02, 0.05, 0.10, 0.15, 0.20, 0.25, 0.30, 0.40, 0.50 \}$. Details about the simulation are introduced in Section \ref{subsec: simulation-microbiome}.} \label{plot: MiRKAT-MiHC}
\end{figure}

%\vspace{-2ex}
Each of the above settings captures different signal structures that may be seen in biological data. We vary the sparsity level for the abundance and random signal settings and effect size for the phylogenetic setting. Details about the data-generating process are presented in Section \ref{subsec: simulation-microbiome}. As seen in Figure \ref{plot: MiRKAT-MiHC}. In the phylogenetic case (a), we see that both methods are similar to one another, but MiHC gains power over MiRKAT at higher effect sizes. In the abundance setting (b), MiRKAT far outperforms MiHC across all sparsity levels while in the random setting (c), MiHC outperforms MiRKAT at low sparsity levels while losing power over MiRKAT as the signal becomes denser. This motivates the need to develop a testing approach that remains robust and powerful under various sparsity levels and signal patterns.

%\vspace{-2ex}
%\vspace{-2ex}

\begin{figure}[!htb]
    \centering
    \includegraphics[width=0.7\textwidth]{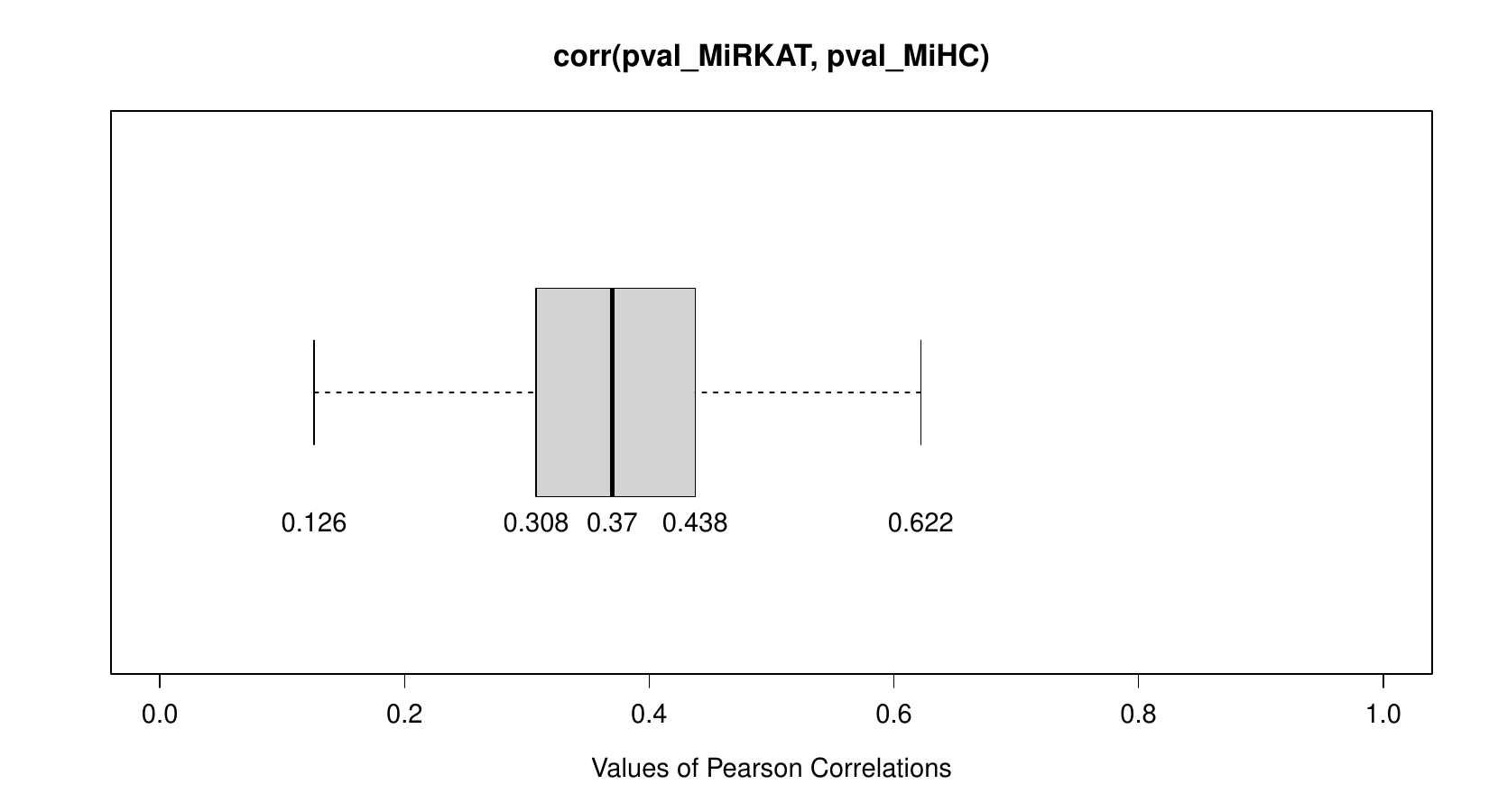}
    \captionsetup{justification=centering}
    %\vspace{-1ex}
    \caption{A Boxplot of the sample Pearson correlations between the $p$-values of MiRKAT and MiHC tests, i.e., corr($p_Q$, $p_M$), under various signal structure settings with 100 replications. The datasets are generated under the null hypothesis.% with $(n, p, p_c, \textrm{MAF}) = (500, 20, 2, 0.1)$. 
    }
    \label{fig: plot-correlation-microbiome}
\end{figure}

To acquire high power over a variety of alternative settings, one natural approach is to base our decision on information from both tests, that is, to construct a new test making use of the two existing tests. We first numerically examine the dependence between MiRKAT and MiHC. 
Figure \ref{fig: plot-correlation-microbiome} shows a boxplot of the sample correlations between the $p$-values of two tests, i.e., $\corr(p_Q, p_M)$. The simulation is conducted under the null hypothesis, and details about the data-generating process are presented in Section 5.1. The plot shows their $p$-values are moderately correlated, which delivers an evident message that the two tests are not independent. Hence, many conventional combination approaches are not readily employable. 

\addtocounter{myalgo}{1}

We utilize the proposed framework \eqref{eq: tildepc} and adapt Algorithm 1 (shown in Table \ref{algo: combining-dependent-$p$-values}) to construct a dependence-adjusted combined microbiome association test.  Detailed procedures are summarized in Algorithm \themyalgo\ (shown in Table \ref{algo: MiRKAT-MiHC}). Theorem \ref{thm: bootstrappedPvalue} can be adapted as well, indicating that the dependence-adjusted test produces a valid combined $p$-value with theoretical guarantees; see Theorem \ref{thm: bootstrappedPvalue-microbiome} below. A proof of the theorem is in Appendix. % \ref{supp-sec: lemmas-proofs}. 

%\vspace{-2ex}
\begin{theorem}\label{thm: bootstrappedPvalue-microbiome}
For a smooth and coordinate-wise monotonic function $g_c$ from $[0,1] \times [0,1] \to \mathbb{R} = (-\infty, \infty)$. Suppose $\bX_i, \bO_i$'s are $i.i.d.$ sub-Gaussian with sub-Gaussian norm bounded by some universal constants respectively. Suppose $(p+q)/n\to 0$, then as $n \to \infty$,
$
\widetilde p_c^* = \widetilde{G}_c^*(g_c(p_Q(\bY), p_M(\bY))) \overset{\cal L}{\to} U[0,1],
$
where $\widetilde G_c^*(s) = \text{\rm P}^*( g_c(p_Q^b, p_M^b) \leq s \mid \bY =  {\by}_{\rm obs})$ for $s \in (-\infty, \infty)$, and $p_Q^b$ and $p_M^b$ are the $p$-values computed using a random bootstrap sample. % Furthermore, 
As $n, B \to \infty$, 
$
\widetilde p_c^B = \widetilde{G}_c^B(g_c(p_Q(\bY), p_M(\bY))) \overset{\cal L}{\to} U[0,1].  
$  %\quad @Linjun: Thm 2?
\end{theorem}

% %\vspace{-3ex}

%%%%%%%%%%%%%%%%%%%%%%%%%%%%%%%%%%%%%%%%%%%%%%%%%%%%%%%%%%%%%%%%%%
%\vspace{-3ex}
\section{Simulation Studies} \label{sec: simulation}
%%%%%%%%%%%%%%%%%%%%%%%%%%%%%%%%%%%%%%%%%%%%%%%%%%%%%%%%%%%%%%%%%%
%In this section, we conduct Monte Carlo simulations to investigate the numerical performance of dependence-adjusted tests constructed through the proposed dependent combination framework. Section \ref{subsec: simulation-microbiome} studies the microbiome association tests. Section \ref{subsec: simulation-dCauchy} considers a special synthetic setting to further elaborate the necessity of adjusting dependence in the combination. 

% ################################################# %
\subsection{Microbiome Association Tests}
\label{subsec: simulation-microbiome}
% ################################################# %
We conduct simulations to examine the numerical performance of our proposed dependence-adjusted combined tests 
in various settings with different sparsity levels and signal structures. Choosing $g_c$ to be $g_c(u_1, u_2) = \log (u_1) + \log (u_2)$, we compare the empirical size and power combined by dependent-Fisher (\texttt{dFisher})  method 
with those by independent-Fisher (\texttt{Fisher}) method.
We also consider different $g_c$ functions listed in Table \ref{tab: gcfuns} and compare their performance in situations of ignoring the dependence 
(denoted by \texttt{Stou}, \texttt{DE}, \texttt{Gcmin}, \texttt{Cauchy})  
and taking dependence into account (denoted by \texttt{dStou}, \texttt{dDE}, \texttt{dGcmin}, \texttt{dCauchy}).

{Following the framework used in \cite{chen2013kernel} and \cite{zhao2015testing}, we generate the OTU information from a Dirichlet-multinomial distribution, which accounts for the potential over-dispersion of OTU count data. To more closely mimic real data, we estimate the dispersion parameter and the proportion means from the study of upper-respiratory microbiome data performed by \cite{charlson2010disordered}, and employ the estimation outputs as the parameter values in the Dirichlet-multinomial distribution for generating simulated OTUs. The upper-respiratory microbiome data from \cite{charlson2010disordered} consists of 856 OTUs. To aid in computational time, we reduce the total number of OTUs to $p=200$. Let $\mathcal{A}$ denote the set of indices for significant OTUs. We study three signal structures (phylogenetic, abundance, and random settings, introduced in Section \ref{subsec: microbiome-bootstrap}) of $\mathcal{A}$. Then, the OTU matrix $\bZ$ is constructed using the OTUs in the signal set $\mathcal{A}$ and the remaining $p-|\mathcal{A}|$ OTUs with the largest abundances, where $|\mathcal{A}|$ denotes the cardinality of set $\mathcal{A}$.} 

Following \cite{zhao2015testing}, 
we generate continuous responses by
%\begin{equation} \label{eq: micro-continuous-size}
    $y_i = 0.5X_{i,1}+0.5X_{i,2}+\epsilon_i$
%\end{equation}
for size evaluation, and
%\begin{equation} \label{eq: micro-continuous-power}
$y_i = 0.5X_{i,1}+0.5X_{i,2}+\mathbf{\beta}\text{scale}(\sum_{j\in \mathcal{A}} Z_{ij})+\epsilon_i$ to evaluate power, $i=1,\ldots, n$. In our numerical studies, we set $n=100$. %{\color{blue} [this formulation seems that you only have three $\beta$'s, so $p = 3$?]}
For each individual, $X_{i,1}$ is a continuous covariate generated from $N(0,1)$, $X_{i,2}$ is a dichotomous covariate following a Bernoulli distribution with success probability 0.5, and $\epsilon_i$ follows $N(0,1)$. 
%\end{equation}
The scale function standardizes the OTU abundance data to mean 0 and standard deviation 1, {and $\beta$ is a scalar that controls the signal strength. To appreciate this, let $\bbeta=(\beta_1, ..., \beta_p)'$ denote the effect size of each OTU. Recall that $\mathcal{A}$ is the index set of significant OTUs. Hence, the effect size of any insignificant OTU (i.e., $j \notin \mathcal{A})$ is $\beta_j = 0$, while the effect size of significant OTUs (i.e., $\beta_j$, $j\in \mathcal{A}$) depends on $\beta$. The sparsity levels among $\{\beta_j, j=1, \ldots, p \}$ are directly determined by the sparsity levels in $\mathcal{A}$ (i.e., the parameter $K$ introduced in Section \ref{subsec: microbiome-bootstrap}). In the phylogenetic setting, we vary the signal strength $\beta \in \{0.2, 0.4, 0.6, 0.8, 1.0\}$. In the abundance and random setting, we fix the signal strength at $\beta=0.5$ and vary the sparsity level $K\% \in \{1\%, 2\%, 5\%, 10\%, 15\%\}$ of $p=200$.}

We present our results averaged over $1,000$ replications in Table \ref{tab: size-mb-continuous} for empirical size. The table shows that ignoring the dependence in the combination procedure may result in a distorted type-I error rate. The \texttt{Fisher}, \texttt{Stouffer}, and \texttt{DE} methods tend to have inflated size, while the \texttt{Gcmin} tends to be conservative. Using bootstrap to account for the dependence, i.e., \texttt{dFisher}, \texttt{dStouffer}, \texttt{dDE}, or \texttt{dGcmin}, makes substantial strides in retaining the desired testing size. We notice that the vanilla \texttt{Cauchy} method exhibits satisfactory performance even though the joint distribution of $(Q_{MiRKAT}, M_{MiHC})$ violates the distributional assumption of bivariate normal. 
The dependence-adjusted Cauchy combination (\texttt{dCauchy}) neither improves nor deteriorates the performance of the \texttt{Cauchy} in this example.

%\vspace{-2ex}
\begin{table}
%\begin{algorithm}
\caption{Microbiome Association Tests: Empirical Size and Power for Continuous Traits}
\begin{subtable}{\textwidth}
\captionsetup{justification=centering}
%\caption{\normalsize Microbiome Association Tests: Empirical Size for Continuous Traits}
%\caption{\normalsize Empirical Size}
\caption{Empirical Size}
\label{tab: size-mb-continuous}
\resizebox{\textwidth}{!}{
\begin{tabular}{r|rrrrrrrrrrrrrrrrr}
 \hline
Sparsity & \multicolumn{1}{c}{MiRKAT} & \multicolumn{1}{c}{MiHC} & \multicolumn{1}{c}{Fisher} & \multicolumn{1}{c}{Stou} & \multicolumn{1}{c}{DE} & \multicolumn{1}{c}{Gcmin} & \multicolumn{1}{c}{Cauchy} & \multicolumn{1}{c}{dFisher} & \multicolumn{1}{c}{dStou} & \multicolumn{1}{c}{dDE} & \multicolumn{1}{c}{dGcmin} & \multicolumn{1}{c}{dCauchy} \\ 
\hline
\multicolumn{12}{c}{Phylogenetic Settings} \\ 
$-$ & 4.4 & 5.4 & 6.5 & 8.3 & 6.7 & 4.1 & 4.9 & 4.5 & 4.4 & 4.5 & 4.6 & 4.3 \\ 
\hline
\multicolumn{12}{c}{Abundance Settings} \\ 
1\%  & 4.5 & 4.8 & 6.1 & 7.1 & 5.8 & 3.8 & 4.5 & 4.8 & 4.7 & 4.7 & 3.9 & 4.1 \\ 
2\%  & 5.8 & 5.9 & 8.1 & 9.5 & 8.5 & 5.7 & 6.8 & 6.9 & 6.8 & 7.0 & 6.1 & 6.4 \\  
5\%  & 4.9 & 5.2 & 6.5 & 7.1 & 6.7 & 4.4 & 5.6 & 5.2 & 5.1 & 5.2 & 4.5 & 4.4 \\ 
10\% & 6.0 & 4.9 & 7.3 & 7.5 & 7.5 & 6.1 & 6.4 & 5.1 & 5.2 & 5.1 & 6.4 & 6.3 \\ 
15\% & 4.5 & 4.3 & 6.7 & 7.7 & 7.0 & 4.2 & 5.0 & 4.5 & 4.8 & 4.6 & 4.6 & 4.3 \\  
\hline
\multicolumn{12}{c}{Random Settings} \\
1\%  & 4.8 & 4.8 & 6.5 & 8.8 & 6.8 & 5.5 & 5.7 & 5.0 & 5.1 & 4.9 & 5.3 & 5.4 \\  
2\%  & 4.2 & 3.9 & 6.4 & 7.4 & 6.6 & 3.3 & 3.8 & 3.9 & 4.6 & 3.9 & 3.0 & 2.8 \\ 
5\%  & 5.6 & 3.7 & 6.1 & 7.0 & 6.3 & 4.7 & 5.3 & 4.5 & 4.4 & 4.5 & 4.6 & 4.5 \\ 
10\% & 5.3 & 5.2 & 5.9 & 7.8 & 6.5 & 4.5 & 4.7 & 5.0 & 4.7 & 4.9 & 5.1 & 4.9 \\ 
15\% & 3.8 & 4.4 & 5.5 & 6.5 & 5.7 & 4.7 & 5.0 & 4.3 & 4.1 & 4.3 & 4.7 & 4.9 \\ 
\hline
\end{tabular}
}
%\noindent\rule{\textwidth}{0.4pt}
\end{subtable}

\bigskip\bigskip
\begin{subtable}{\textwidth}
\small
\captionsetup{justification=centering}
%\caption{\normalsize Microbiome Association Tests: Empirical Power for Continuous Traits} 
%\caption{\normalsize Empirical Power}  \label{tab: power-mb-continuous}
\caption{Empirical Power}\label{tab: power-mb-continuous}
\centering
\label{algo: MiRKAT-MiHC}
%\noindent\rule{\textwidth}{0.4pt}
\resizebox{\textwidth}{!}{
\begin{tabular}{rr|rrrrrrrrrrrrrrrrr}
 \hline
Sparsity & Eff.Size & \multicolumn{1}{c}{MiRKAT} & \multicolumn{1}{c}{MiHC} & \multicolumn{1}{c}{dFisher} & \multicolumn{1}{c}{dStou} & \multicolumn{1}{c}{dDE} & \multicolumn{1}{c}{dGcmin} & \multicolumn{1}{c}{dCauchy} \\
\hline
 & & \multicolumn{7}{c}{Phylogenetic Settings} \\ 
$-$ & 0.2 &  7.6 &  6.6 &  7.0 &  6.9 &  7.1 &  7.1 &  7.0 \\ 
$-$ & 0.4 & 14.2 & 16.0 & 19.3 & 19.3 & 19.2 & 16.6 & 17.2 \\ 
$-$ & 0.6 & 29.8 & 36.5 & 42.3 & 43.3 & 42.3 & 37.8 & 39.4 \\  
$-$ & 0.8 & 52.3 & 67.6 & 72.4 & 72.4 & 72.5 & 67.8 & 69.6 \\ 
$-$ & 1.0 & 63.5 & 85.4 & 89.1 & 89.3 & 89.1 & 84.0 & 85.9 \\ 
\hline
 & & \multicolumn{7}{c}{Abundance Settings} \\ 
1\%  & 0.5 & 89.6 & 30.8 & 81.1 & 73.7 & 79.5 & 85.4 & 84.9 \\ 
2\%  & 0.5 & 86.5 & 14.0 & 74.9 & 63.8 & 73.3 & 79.7 & 79.0 \\ 
5\%  & 0.5 & 82.8 &  8.1 & 68.4 & 53.6 & 66.3 & 75.4 & 74.0 \\ 
10\% & 0.5 & 78.6 &  6.8 & 60.1 & 45.1 & 57.7 & 70.4 & 68.8 \\
15\% & 0.5 & 75.7 &  5.2 & 55.4 & 40.2 & 52.5 & 65.8 & 63.4 \\ 
\hline
 & & \multicolumn{7}{c}{Random Settings} \\
1\%  & 0.5 &  8.8 & 73.0 & 61.4 & 48.8 & 58.9 & 65.6 & 64.3 \\ 
2\%  & 0.5 & 12.1 & 61.1 & 49.4 & 43.5 & 48.5 & 54.1 & 53.4 \\  
5\%  & 0.5 & 18.6 & 45.8 & 41.8 & 38.4 & 41.3 & 41.8 & 42.3 \\ 
10\% & 0.5 & 22.6 & 34.7 & 35.5 & 35.7 & 35.4 & 34.4 & 34.8 \\ 
15\% & 0.5 & 27.4 & 27.9 & 34.7 & 32.8 & 34.8 & 34.0 & 35.2 \\ 
\hline
\end{tabular}
}
%\noindent\rule{\textwidth}{0.4pt}

 \vspace{2ex}
{
\centering
Note: This table reports the frequencies of rejection by each method under the alternative hypothesis based on 1000 independent replications conducted at the significance level $5\%$. The dependent combination methods (e.g., dFisher, dStou, dDE, dGcmin, dCauchy) are implemented using $B=500$. 
}
\end{subtable}
\end{table}
%\vspace{-2ex}

%\vspace{-2ex}
As discussed above, ignoring the dependence can lead to distorted size. When assessing the power, our focus solely rests on the dependence-adjusted methods. A comparison of empirical power is presented in Table \ref{tab: power-mb-continuous}. It can be observed that the combined test methods are more robust than MiRKAT and MiHC in various sparsity levels and signal structures. For example, MiRKAT tends to outperform MiHC in the abundance setting, while this scenario flips in the random setting. Our proposed dependence-adjusted combined tests attain consistent power regardless of the signal pattern. In practice, the true signal pattern may not be well understood and an incorrect choice of the test may lead to a large loss of discovery power. By employing our proposed dependence-adjusted combination tests, we are able to ensure competitive power across a variety of alternative settings.

% ############################################################## %
%\vspace{-2.5ex}
\subsection{Additional Simulation Studies on the Cauchy Combination Method}
\label{subsec: simulation-dCauchy}
% ############################################################## %
In the sequel, we study a special synthetic setting to further elaborate on the necessity of adjusting dependence in the combination. Though the numerical results in Sections \ref{subsec: simulation-microbiome} suggest that the vanilla Cauchy combination test may perform reasonably well even with its validity not theoretically guaranteed due to violations of assumptions, it is not always the case. In what follows, we conduct additional studies on the Cauchy combination to investigate its performance for combining arbitrarily dependent components. We discover a synthetic setting in which the vanilla Cauchy combination (denoted by \texttt{Cauchy}) fails to control the type-I error. In contrast, the dependence-adjusted Cauchy (denoted by \texttt{dCauchy}) retains the correct size after accommodating the dependence. 

We consider the following synthetic setting: 
$X_1$ and $X_2$ are independent, and they follow the standard Cauchy distribution. Let 
%\vspace{-2ex}
\begin{align}
    T_1 = X_1 \quad \text{and}\quad 
    T_2 = \left\{
    \begin{aligned}
    |X_2|, \quad T_1 \geq 0, \\
    -|X_2|, \quad T_1 < 0.
    \end{aligned}
    \right.
    %\vspace{-2ex}
\end{align}
Both $T_1$ and $T_2$ are marginally standard Cauchy variables, but their joint distribution is not bivariate Cauchy. 
We treat $T_1$ and $T_2$ as two test statistics, and define $p_1 = 1-F_{Cauchy}(T_1)$ and $p_2 = 1-F_{Cauchy}(T_2)$. Then $p_1$ and $p_2$ can be regarded as two $p$-values associated with test statistics $T_1$ and $T_2$. Both $p_1$ and $p_2$ follow $\mathcal{U}[0,1]$ marginally, and $p_1$ and $p_2$ are not independent. Define 
$T_{Cauchy} = 0.5 \tan \{ (0.5- p_1) \pi \}  + 0.5 \tan \{ (0.5- p_2) \pi \}. $

The Cauchy combination approximates the tail probability of $T_{Cauchy}$ using the standard Cauchy distribution and computes the $p$-value  as $p_{Cauchy} = 1-F_{Cauchy}(T_{Cauchy})$. If $p_1$ and $p_2$ are independent, then $T_{Cauchy}$ is naturally a standard Cauchy variable. \cite{liu2020Cauchy} proved  the tail behavior of $T_{Cauchy}$ could be well approximated by a Cauchy distribution if $(T_1, T_2)$ follows a bivariate normal distribution. Such bivariate normal assumption accommodates the dependency to some extent but has not yet been generalized to arbitrary joint distributions. By construction, it is straightforward to see the bivariate normal assumption does not hold in this example. In fact, \cite{lee2014clarification} proved for this example $T_{Cauchy}$ is not Cauchy distributed.  The distribution of $T_{Cauchy}$ remains an open question for arbitrarily distributed $(T_1, T_2)$. As a result, the Cauchy combination should not be blindly applied here. 

%\vspace{-2ex}
\begin{figure}%[!htb]
\centering
\includegraphics[scale=0.4]{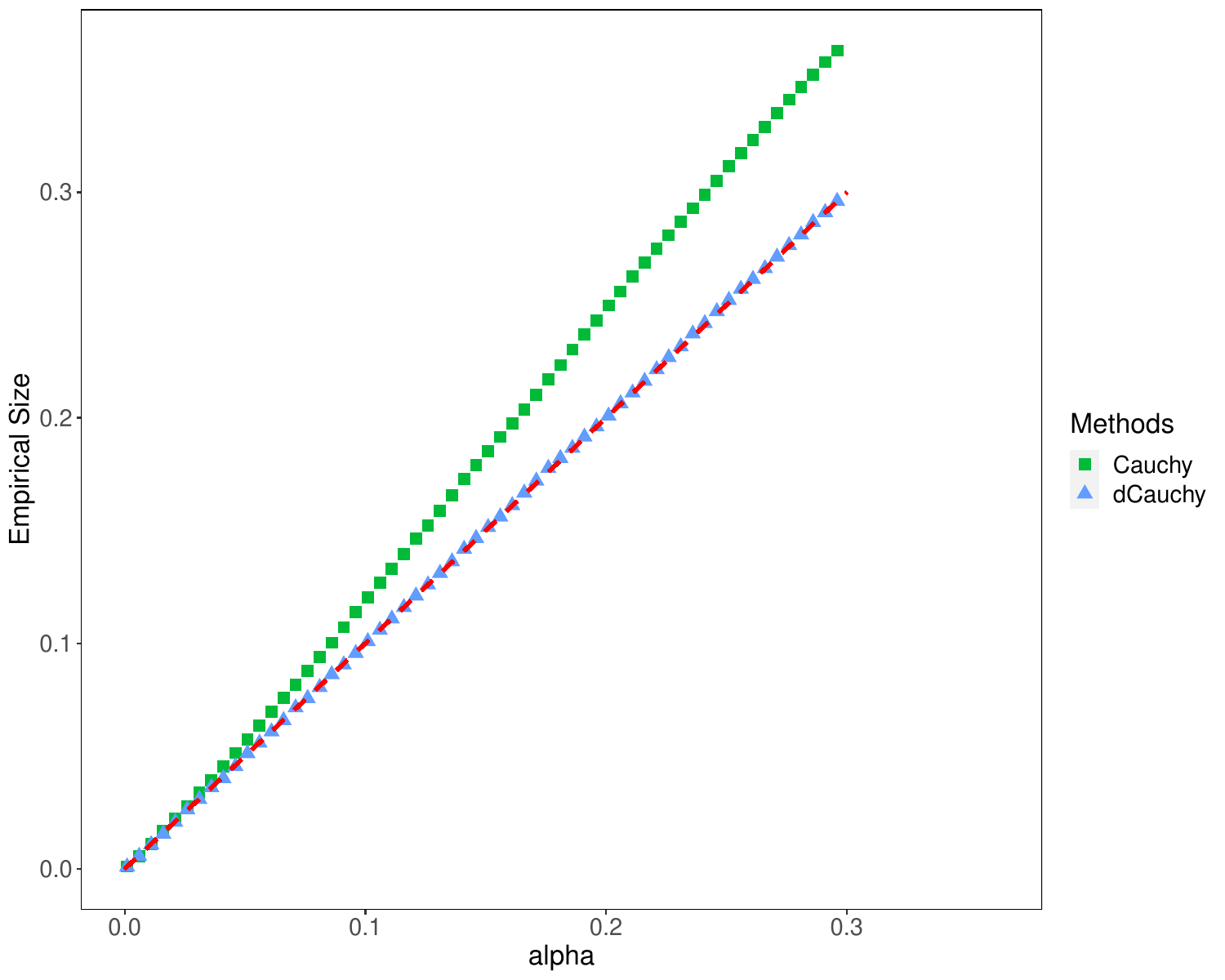}
%%\vspace{-1ex}
\caption{Results of empirical size over 100,000 replications. 
\texttt{dCauchy} is implemented via model-based bootstrap with $B=2000$. The red line indicates values of the nominal level. } \label{Fig: Cauchy-dCauchy-size}
\end{figure}
%\vspace{-2ex}

We compare the performance of \texttt{Cauchy} and \texttt{dCauchy} in which the impact brought by the dependence is adjusted using our proposed framework. Specifically, we investigate their performance in controlling Type-I error rate. The results are displayed in Figure \ref{Fig: Cauchy-dCauchy-size} based on 100,000 replications with the significance level $\alpha$ varying from 0.001 to 0.3 as a multiplier of 0.005. %The red line indicates the values of nominal level $\alpha$.
The plot shows that in this example, the vanilla Cauchy combination fails to provide a good approximation of the tail probability of $T_{Cauchy}$. The {\texttt{Cauchy}} method yields inflated Type-I errors due to negligence of the dependence. Using the proposed framework, the {\texttt{dCauchy}} %takes the dependence into account and 
offers accurate size approximation. The observations from this synthetic example further emphasize the importance of taking dependence into account.

%%%%%%%%%%%%%%%%%%%%%%%%%%%%%%%%%%%%%%%%%%%%%%%%%%%%%
%\vspace{-3ex}
\section{Real Data Examples} \label{sec: realdata}
%%%%%%%%%%%%%%%%%%%%%%%%%%%%%%%%%%%%%%%%%%%%%%%%%%%%%

We demonstrate the effectiveness of our proposed methods on two real-world datasets. The first analysis focuses on determining the association between the respiratory microbiome and the smoking status of individuals while the second analysis focuses on the association between the gastrointestinal microbiome and the development of inflammatory bowel disease.

We employ the MiRKAT and MiHC association tests and also demonstrate the flexibility of our combination test framework by studying the weighted Microbiome Sum of Powered Score test (MiSPU) \citep{wu16adaptive}. The MiSPU was proposed to test the association between microbial features and outcome of interest through score statistics derived from the underlying data model. The MiSPU method requires the selection of a tuning parameter $\gamma$ which controls the weight placed on individual score components. Notably, as $\gamma$ increases, larger score components are weighted more heavily. For MiSPU($\gamma$), we vary $\gamma \in \{2,3,4,5,6,7,8\}$ and use MiSPU(5) in our combination tests. We run the dependence-adjusted combination tests with $B=500$ bootstrap iterations. For both analyses, we target an $\alpha = 0.05$ level.

%\subsection{Smoking Data Analysis}
\smallskip
\noindent \textbf{\textit{(a) Smoking Data Analysis. }}
We study the respiratory microbiome data analyzed by \cite{charlson2010disordered}, which can be accessed as $\texttt{throat}$ data in \texttt{GUniFrac} R package. This dataset contains the microbiome information of $n=60$ individuals including 32 non-smokers and 28 smokers. Thus, the goal of this analysis is to determine if the microbiome community is associated with the smoking status of a subject. We first pre-process the raw OTU data to remove singleton and monotone OTUs. Further, we retain the $p=103$ OTUs with mean relative abundance values greater than $10^{-3}$. Due to the nature of the data, we also control for clinical covariates of gender and if the individual had used antibiotics in the last 3 months.

%\vspace{-2ex}
\begin{table}%[ht]
\captionsetup{justification=centering}
%\caption{The computed $p$-values for weighted MiSPU using various choices of weight parameters $\gamma$ for the Smoking and the Inflammatory Bowel Disease datasets.}
\caption{The computed $p$-values of various testing procedures for the Smoking and the Inflammatory Bowel Disease datasets.}
\begin{subtable}{\textwidth}
\captionsetup{justification=centering}
%\caption{Computed $p$-values for weighted MiSPU using various choices of weight parameters $\gamma$ for the Smoking dataset.}\label{SPU-smoke}
\caption{\normalsize The Smoking Dataset.}\label{SPU-smoke}
\centering
\begin{tabular}{crrrrrrrrrr}
  \hline
Methods  & $p$-values \\ \hline 
MiRKAT   & 0.003 \\
MiHC     & 0.018  \\ 
MiSPU(2) & 0.106 \\
MiSPU(3) & 0.291 \\ 
MiSPU(4) & 0.284 \\ 
MiSPU(5) & 0.295 \\ 
MiSPU(6) & 0.296 \\ 
MiSPU(7) & 0.297 \\ 
MiSPU(8) & 0.297 \\  
MiRKAT+MiHC (dFisher) & 0.002 \\
MiRKAT+MiHC (dCauchy) & 0.000 \\
MiSPU(5)+MiHC (dFisher) & 0.024 \\ 
MiSPU(5)+MiHC (dCauchy) & 0.012 \\ 
   \hline
\end{tabular}
\end{subtable}

\bigskip\bigskip\bigskip
\begin{subtable}{\textwidth}
\captionsetup{justification=centering}
%\caption{Computed $p$-values for weighted MiSPU using various choices of weight parameters $\gamma$ for the Inflammatory Bowel Disease dataset.}\label{SPU-CCFA}
\caption{\normalsize The Inflammatory Bowel Disease Dataset. \\
}\label{SPU-CCFA}
\centering
\begin{tabular}{crrrrrrrr}
  \hline
Methods  & $p$-values \\ \hline
MiRKAT   & 0.011 \\
MiHC     & 0.210  \\ 
MiSPU(2) & 0.008 \\ 
MiSPU(3) & 0.006 \\ 
MiSPU(4) & 0.008 \\ 
MiSPU(5) & 0.008 \\ 
MiSPU(6) & 0.007 \\ 
MiSPU(7) & 0.007  \\ 
MiSPU(8) & 0.007 \\ 
MiRKAT+MiHC (dFisher) & 0.020 \\
MiRKAT+MiHC (dCauchy) & 0.022 \\
MiSPU(5)+MiHC (dFisher) & 0.008 \\ 
MiSPU(5)+MiHC (dCauchy) & 0.012 \\ 
   \hline
\end{tabular}

 %\vspace{5ex}
   {
   %\small
      Note: The last four rows report the combined $p$-values from different tests using various combination approaches. For example, MiRKAT+MiHC (dFisher) denotes the combined $p$-value from MiRKAT and MiHC using the dFisher approach.  The dependent combination methods are implemented using $B=500$. }
      
\end{subtable}

\end{table}
%\vspace{-2ex}

The results of various methods are summarized in Table \ref{SPU-smoke}. 
Firstly, at all levels of $\gamma$, 
MiSPU does not find a significant association between the respiratory microbiome and smoking status. However, MiRKAT identifies a significant association with $p_{MiRKAT} = 0.003$. Likewise, MiHC identified a significant association with $p_{MiHC} = 0.018$. When comparing the dependence-adjusted tests, the combination of MiRKAT and MiHC is also significant regardless of the combination test with $p^{dFisher}_{MiRKAT+MiHC} = 0.002$ and $p^{dCauchy}_{MiRKAT+MiHC} = 0$. Likewise the combination of MiSP(5) and MiHC $p^{dFisher}_{MiSPU(5)+MiHC} = 0.024$ and $p^{dCauchy}_{MiSPU(5)+MiHC} = 0.012$. As noted previously, a key benefit of our proposed combination test procedure is the combination tests are more robust in settings where the underlying signal sparsity may not be known. While MiSPU may not have identified a significant relationship, all other methods show a strong significance between the respiratory microbiome and smoking status.

This significant association has been noted in the literature as well. Due to the sensitivity of the lung microbiome, smoking can drastically disrupt the microbial environment. Smoking has been linked to changes in the diversity of the lung microbiome which may contribute to the development of inflammatory disease. Further, due to the complex chemical makeup of cigarette smoke, it has been shown that this may affect the severity of bacterial infections and risks of chronic lung disease \citep{garmendia2012impact}.

%\subsection{Inflammatory Bowel Disease Data}
%\smallskip
\noindent \textbf{\textit{(b) Inflammatory Bowel Disease Data. }} We study the gut microbiome data analyzed by \cite{gevers2014treatment}. Chron's disease and other inflammatory bowel diseases are common afflictions that have been noted to be linked to changes in the gut microbiome. This analysis tests for the association between the composition of the gut microbiome and whether an individual has developed inflammatory bowel disease. This data can be obtained from the \texttt{RISK CCFA} dataset as part of the \texttt{MicrobeDS} R-package. %\citep{microbeds}. 
This data set is expansive, containing information of $p_0 = 9511$ OTUs from $n_0 = 1359$ individuals. After removing the control samples, we have 1023 individuals who suffer from gastrointestinal disease and 355 who do not. Due to the scale of this dataset, we focus on the top $p=300$ OTUs with the highest abundances and randomly sub-sample $n=150$ individuals from this data set. We control for variation attributed to an individual's race and gender.

The results of different association tests are presented in Table \ref{SPU-CCFA}. 
We find that the MiRKAT method yields a significant association with a $p$-value of $p_{MiRKAT} = 0.011$. For this data, we find the MiHC method does not find a significant association with $p_{MiHC} = 0.210$. However, at all weights $\gamma$, we find that MiSPU($\gamma$) identifies a significant association. % as seen in Table \ref{SPU-CCFA}.
When comparing the dependence-adjusted combination tests. the combination of MiRKAT and MiHC is significant with $p^{dFisher}_{MiRKAT+MiHC} = 0.020$ and $p^{dCauchy}_{MiRKAT+MiHC} = 0.022$. The combination of MiSPU and MiHC is also significant with $p^{dFisher}_{MiSPU(5)+MiHC}=0.008$ and $p^{dCauchy}_{MiSPU(5)+MiHC} = 0.012$. A majority of these tests suggest a significant association between the gut microbiome and the risk of gastrointestinal disease. Notably, this echoes the findings in biological literature. The work by \cite{gevers2014treatment} finds significant links between the two and suggests that early intervention to alter the gut microbiome can reduce the onset and severity of Chron's disease. In a similar manner, \cite{morgan2012dysfunction} derived similar conclusions by noting the changes in the abundance of key bacteria within the gut microbiome can greatly influence the severity of various types of inflammatory bowel disease.

%\vspace{-4ex}

%%%%%%%%%%%%%%%%%%%%%%%%%%%%%%%%%%%%%%%%%%%%%%%%%%%%%%%%%%%%%%%%%%

\section{Efficiency of Various Combination Methods} \label{sec: various-combination-methods}
%%%%%%%%%%%%%%%%%%%%%%%%%%%%%%%%%%%%%%%%%%%%%%%%%%%%%%%%%%%%%%%%%%
% ############################################################## %
\subsection{Combination of Dependent Normal Statistics}\label{subsec: dep-stat}
% ############################################################## %

We use a simple example to begin our investigation of the efficiency of different combination methods. Assume  {$Z_1 \sim N(\mu, 1/n)$}, {$Z_2 \sim N(\mu, 1/n)$} are bivariate normal with a (for simplicity, known) correlation  $\rho = \corr(Z_1, Z_2)$. Consider a one-sided test
\vspace{-2ex}
\begin{equation} \label{eq: hypo-one-sided}
    H_0^{(1)}: \mu \leq \mu_0 \quad \hbox{versus} \quad H_1^{(1)}: \mu > \mu_0.
    \vspace{-2ex}
\end{equation}
The $p$-values are obtained by $p_i = p_i(\mu_0)= \Phi\left(\sqrt{n}(Z_i - \mu_0)\right)$, $i = 1,2$. The two $p$-values $p_1$ and $p_2$ may not be independent when $\rho\neq 0$, and the dependence between the two $p$-values can be explicitly computed using a Gaussian copula. 

The most powerful test is associated with the average of $Z_1$ and $Z_2$, i.e., $\bar{Z} = \frac{1}{2}(Z_1 + Z_2) \sim N(\mu, \frac{1}{2n}(1+\rho))$. The rejection region at the significance level $\alpha$ is { $\{{\bar{z}}: {\bar{z}} - \mu_0 \geq z_{\alpha}\sqrt{(1+\rho)/2n}\}$}. The power function at $\mu = \mu_1 > \mu_0$ is
\vspace{-1ex}
\begin{equation} \label{eq: power-fun-MP-dependent}
    A_{MP}^{d}(\mu_1) = \PP^d_{\mu_1}(\bar{Z} - \mu_0 > z_{\alpha}\sqrt{(1+\rho)/2n} ) = 1 - \Phi(z_{\alpha} - \sqrt{2n/(1+\rho)}(\mu_1 - \mu_0)
),
\vspace{-1ex}
\end{equation}
Here ``MP'' stands for Most Powerful, and $\PP^d$ emphasizes the possible dependence between $Z_1$ and $Z_2$. Stouffer's method combines two tests by taking $g_c(u_1, u_2) = \Phi^{-1}(u_1) + \Phi^{-1}(u_2)$. With $(U_1, U_2)$ from the Gaussian copula of correlation $\rho$, we have $\widetilde G_c(s) = P(g_c(U_1, U_2)\leq s) = \Phi(s/\sqrt{2(1+\rho)})$. The combined $p$-value is obtained by $\widetilde{p}_c(\mu_0)
=  \Phi( (\sqrt{n}(Z_1 - \mu_0) + \sqrt{n}(Z_2 - \mu_0))/\sqrt{2(1+\rho)} ) 
= \Phi( \sqrt{2n}(\bar{Z} - \mu_0)/\sqrt{1+\rho} ),
$ The test rejects $H_0$ at level $\alpha$ if $\bar{Z} - \mu_0 \geq z_{\alpha}\sqrt{(1+\rho)/2n}$, which coincides with the rejection region of the MP test. This leads to the same power function as that of the MP test \eqref{eq: power-fun-MP-dependent}. In other words, Stouffer's method produces the most powerful test when combining $p$-values from {this pair of} $Z$-tests.

The conventional Fisher's method \citep{Fisher1925} relies on the assumption of independence among different components. 
However, neglecting dependence often leads to severe size distortion, and hence adjustments for dependence are necessary for the combination of dependent $p$-values, for instance, the $p_1$ and $p_2$ in this example. The dependence-adjusted Fisher's method uses the same function $g_c(u_1, u_2) = \log (u_1) + \log (u_2)$ as in the independent case but, in the definition of $\widetilde G_c(s)$, $(U_1, U_2)$ are no longer independent. 
This $\widetilde G_c(s)$ does not have a closed-form expression; thus neither does the power function (denoted by $A_{FC}^d(\mu_1)$) of the dependence-adjusted Fisher's method.

The recently proposed Cauchy combination by \cite{liu2020Cauchy} has brought much attention 
to the combination of dependent tests. As we discussed previously, the method by \cite{liu2020Cauchy} is a special case of the independent combination framework \eqref{eq:ind} with $g_c(u_1, u_2) = 0.5 F_{CC}^{-1}(u_1)+ 0.5 F_{CC}^{-1}(u_1)$, where $F_{CC}(t) = \frac{1}{2} + \frac{1}{\pi} \textrm{arctan}(t)$ and $F_{CC}^{-1}(u) = \tan [(u - 0.5) \pi]$ are the CDF and the inverse CDF of the standard Cauchy distribution respectively. If $p_1$ and $p_2$ are independent, then $G_c(t)$ $= F_{CC}(t)$ and the associated $p$-value is
$
p_{CC}(\mu_0) = F_{CC}\big(
\frac 12 F_{CC}^{-1}\big(\Phi\big(\sqrt{n}(Z_1 - \mu_0)\big)\big) + \frac12 F_{CC}^{-1}\big(\Phi\big(\sqrt{n}(Z_2 - \mu_0)\big)\big)
\big). $ The corresponding rejection region at significance level $\alpha$ is 
$\big\{ (z_1, z_2): 0.5 F_{CC}^{-1}\big(\Phi\big(\sqrt{n}(z_1 - \mu)\big)\big) + 0.5  F_{CC}^{-1}\big(\Phi\big(\sqrt{n}(z_2 - \mu)\big)\big) > c_{\alpha} \big\}, 
$
where $c_{\alpha}$ is the upper-$\alpha$ quantile of the standard Cauchy distribution. Its {power at $\mu = \mu_1 > \mu_0$ is
%\begin{equation*} 
$A_{CC}(\mu_1) =  \PP_{\mu_1}({ 0.5F_{CC}^{-1}(\Phi(\sqrt{n}(Z_1 - \mu_0))) + 0.5F_{CC}^{-1}(\Phi(\sqrt{n}(Z_2 - \mu_0)))} > c_{\alpha} ).$ 
%\end{equation*}
%Here ``CC'' stands for Cauchy Combination. 
For dependent $p_1$ and $p_2$, when using the method of \cite{liu2020Cauchy}, the level-$\alpha$ rejection region is exactly the same as that of the independent case. Its power at $\mu = \mu_1 > \mu_0$ is, however, 
%\begin{equation*} %\label{eq: Cauchy-power}
$A^d_{CC}(\mu_1) =  \PP_{\mu_1}^d({0.5F_{CC}^{-1}(\Phi(\sqrt{n}(Z_1 - \mu_0))) + 0.5 F_{CC}^{-1}(\Phi(\sqrt{n}(Z_2 - \mu_0)))} > c_{\alpha} ).$ 
%\end{equation*}

The power functions of both the dependence-adjusted Fisher combination and the Cauchy combination do not have explicit formulas, whereas we know that $A^d_{FC}(\mu_1)< A_{MP}(\mu_1)$ and $A^d_{CC}(\mu_1) < A_{MP}(\mu_1)$ since $A^d_{MP}(\mu_1)$ is the most powerful.

% ############################################################## %
%\vspace{-3ex}
\subsection{A Summary and Discussion on the Combination Efficiency} \label{subsec: discussion-efficiency}
% ############################################################## %

The examples above consider the cases when statistics $Z_1$ and $Z_2$ share equal variance. Hence, Stouffer's combination method posits equal weights on the $p$-values resulting from the two statistics, which produce exactly the same rejection regions and power analysis as those of the most powerful test. When the variances of $Z_1$ and $Z_2$ are not equal, the weights need to be appropriately adjusted so Stouffer's combination test can coincide with the most powerful test. 
These discussions show that Stouffer's method with appropriate weights induces the most powerful test when combining $p$-values of normally distributed test statistics, implying that Stouffer's combination is the most efficient for the combination of $p$-values obtained from $Z$-tests. It is worth mentioning that the most efficient claim of Stouffer's method is only limited to $Z$-tests (i.e., normal cases). When the statistics are not normal variables, there is no guarantee that Stouffer's combination test coincides with the most powerful test. 

Fisher's method (or the Double-Exponential combination) is the most Bahadur efficient for combining $p$-values from independent tests for testing one-sided (or two-sided) hypotheses. \cite{littell1971asymptotic,littell1973asymptotic} proved Fisher's method is optimal with respect to Bahadur efficiency \citep{bahadur1967rates} when the $p$-values take the form of $p_i = 1-F_i(Z_{i, obs})$ with $F_i(z) = \PP_{H_0}(Z_i \leq z)$ and $Z_{i, obs}$ being the null CDF and the observed value of test statistics $Z_i$, which is associated with one-sided tests.   
Among all combined $p$-values, the negative log $p$-value obtained by Fisher's combination method decays the fastest under the alternative.
\cite{singh2005combining} and \cite{xie2011confidence} showed the Double-Exponential combination achieves Bahadur optimality for two-sided tests. The Bahadur efficiency results do not require distributional assumptions on test statistics. Specifically, the statement holds for the combination of $Z$-tests in which the test statistics are normally distributed and for all types of tests when $F_i(z)$ can be any valid CDF.

Both Stouffer's and Fisher's combination methods operate under the assumption of independence among the $p$-values. The approaches can be extended to dependent scenarios, provided that the effects of dependence are appropriately taken into account. Especially when the dependence is unknown, resampling methods, such as bootstrapping, may be needed to estimate the dependence. %In Section \ref{sec: methodology}, we propose a unifying dependent combination framework that includes both methods as special cases. 
In contrast, the Cauchy combination can accommodate the dependence to some extent, specifically in the situation when the statistics follow a bivariate normal distribution \citep{liu2020Cauchy}. Some empirical studies have shown that the Cauchy combination may still be valid under a more relaxed dependence structure even when the bivariate normal assumption is violated \citep{liu2019acat}. However, the theoretical guarantee of the Cauchy combination under more relaxed assumptions remains unclear unless our extended approach of (\ref{eq: tildepc}) is used. As a matter of fact, Section \ref{subsec: simulation-dCauchy} studies a synthetic setting when the Cauchy combination test fails to control the Type-I error rate, indicating that the Cauchy approximation cannot be blindly applied to arbitrarily dependent situations under arbitrary distributional assumptions. As for efficiency, the discussions in the previous two subsections show that for the combination of independent $Z$-tests, the Cauchy combination is neither the most powerful nor the one with Bahadur efficiency. That said, the validity and efficiency of the Cauchy combination test by \cite{liu2020Cauchy} under general distributional assumptions and dependence structure remain open questions.

%%%%%%%%%%%%%%%%%%%%%%%%%%%%%%%%%%%%%%%%%%%%%%%%%%%%%%%%%%%%%%%%%%
%\vspace{-5ex}
\section{Conclusion and Discussions} \label{sec: conclusion}
%%%%%%%%%%%%%%%%%%%%%%%%%%%%%%%%%%%%%%%%%%%%%%%%%%%%%%%%%%%%%%%%%%

This paper introduces a unifying combination framework to aggregate information by combining multiple $p$-values. The proposed framework extends \cite{xie2011confidence} of combining independent studies and allows for the existence of dependence. It connects to many conventional meta-analysis methods %, e.g., Fisher's method, Stouffer's method, and Tippett's method, 
when the $p$-values are independent, and provides adjustments to take account of the dependence when the to-be-combined components are dependent. The proposed method provides statistical guarantees for the resulting dependence-adjusted combined tests under arbitrary distributional assumptions of the test statistics and arbitrarily dependent structure among the $p$-values.  We comprehensively study various $p$-value combination methods under the framework and discuss the efficiency of different combination methods. Using microbiome association tests as an application and an illustrative example, we show that the combined tests achieve robust and enhanced power across a wide range of alternatives. Both simulation studies and real applications demonstrate the effectiveness of the combined tests in controlling Type-I error rate, maintaining high power, and achieving test robustness with respect to various alternative hypotheses.   

Recently, researchers have made continuous progress in the combination of dependent $p$-values, e.g., \cite{ wilson2019harmonic,liu2020Cauchy}, and others. Our proposed framework is general and inclusive by incorporating many popular dependent and independent combination methods %In addition to the examples listed in Table \ref{tab: gcfuns}, our proposed framework can also make connections to other combination methods, such as the harmonic mean $p$-value approach \citep{wilson2019harmonic}, 
by letting the $g_c(\cdot)$ function take various forms. Using the $g_c(\cdot)$ function greatly increases the generality and flexibility of our proposed framework. It diversifies the combination approaches this framework can admit and brings flexibility by imposing different weights on various components. If we have some prior knowledge that the information from one aspect should be more important than the other, then we can assign more weight to the important one. Though the examples in Table \ref{tab: gcfuns} assign equal weights to all the components, we can easily modify the $g_c(\cdot)$ by adding weight coefficients. 

Moreover, our proposed framework improves over existing dependent combination methods by providing valid $p$-values under arbitrary distributional assumptions and arbitrarily dependent structures. Most existing dependent combination approaches in the literature adopt one of the following two strategies to handle dependency. (i) Some methods design their decision rules to protect against the worst possible scenario, typically making them overly conservative and lacking power. Examples include the Bonferroni method, the average $p$-value method \citep{ruschendorf1982random},  (ii) Other approaches approximate the combined $p$-values under some additional distributional assumptions, such as the vanilla Cauchy combination \citep{liu2019acat,liu2020Cauchy}, the harmonic mean method \citep{wilson2019harmonic}, the heavy-tailed transformation \citep{fang2023heavy}. %{\cite{fang2023heavy} further extends the works of \cite{liu2020Cauchy} and \cite{wilson2019harmonic} using a broad family of heavy-tailed distributions.} 
These methods only provide approximation under certain conditions but fail to produce valid $p$-values. %; % see Remark \ref{remark:Cauchy}. %and Remark \ref{remark:heavyTail}.
In contrast, our proposed framework explicitly takes into account the dependence via parametric bootstrap, making our combined tests accurate and powerful.

%%\vspace{-4ex}
%\section*{Acknowledgements}
%%\vspace{-2ex}
%The research is supported in part by NIH grant 1R01GM152812, and NSF grants DMS-1811552 \& DMS-1953189.
%

%\setstretch{1.2}
\bibliographystyle{apalike}
\bibliography{association.bib}

\end{document}